\documentclass{aipproc}

\makeatletter

\def\nhline{\noalign{\ifnum0=`}\fi\hrule \@height \arrayrulewidth 
\futurelet
   \@tempa\@xhline}

\def\phantomthreeptline{\noalign{\ifnum0=`}\fi\vskip 3pt 
\futurelet
   \@tempa\@xhline}

\def\make@abstract{{\normalsize\typeout{\abstractname}
  {\noindent{\bf \abstractname}\@abstract}\gdef\@abstract{}
  \if@keywords{%
    \vskip 1\baselineskip\typeout{\keywordsname}
    \noindent{\bf \keywordsname:\/} \@keywords\gdef
    \@keywords{}%
}\fi
\par}}

\makeatother

\def\im{{\rm i}}
\def\ie{{\it i.e.}}
\def\eg{{\it e.g.}}
\def\etc{{\it etc}}
\def\dalemb#1#2{{\vbox{\hrule height .#2pt
        \hbox{\vrule width.#2pt height#1pt \kern#1 pt
                \vrule width.#2 pt}
        \hrule height.#2 pt}}}
\def\square{\mathord{\dalemb{6.2}{6}\hbox{\hskip1pt}}}

\newcommand{\be}{\begin{equation}}
\newcommand{\ee}{\end{equation}}
\newcommand{\bea}{\begin{eqnarray}}
\newcommand{\eea}{\end{eqnarray}}

\newcommand{\ssst}{\scriptscriptstyle}
\def\ft#1#2{{\textstyle{{\scriptstyle #1}\over {\scriptstyle #2}}}}

\layoutstyle{6x9}

\SetInternalRegister\hbadness{8000}

\newcommand\doingARLO[2][]{%
  \ifx\mmref\undefined #1\else #2\fi
}

\begin{document}
\begin{flushright}
\hfill{Imperial/TP/1-02/17}\\
\hfill{hep-th/0203015}
\end{flushright}

\title{Revisiting Supergravity and Super Yang-Mills Renormalization}

\author{K.S. Stelle}{
  address={Department of Physics, Imperial College, London SW7 2BW, UK},
  email={k.stelle@ic.ac.uk},
}

\begin{abstract}
Standard superspace Feynman diagram rules give one estimate of the onset of
ultraviolet divergences in supergravity and super Yang-Mills theories. Newer
techniques motivated by string theory but which also make essential use of
unitarity cutting rules give another in certain cases. We trace the difference
to the treatment of higher-dimensional gauge invariance in supersymmetric
theories that can be dimensionally oxidized to pure supersymmetric gauge
theories.
\end{abstract}


\maketitle

\section{Non-renormalization theorems}

The problem we shall consider in this article\footnote{This article is based
upon work done in collaboration with P.S. Howe and M. Petrini.} is to reconcile
the predictions for the onset of ultraviolet divergences in supergravity and
super Yang-Mills theories
\cite{stelle:howestellerev}made from traditional superspace Feynman diagram
analysis and those made on the basis of unitarity and factorization implied by
embedding in string theory
\cite{stelle:bddpr}. The question of ultraviolet divergences is one of the
oldest concerns in the subject of quantum gravity. Although the problem of
order-by-order perturbative finiteness has been solved in superstring theory,
the question remains of interest because superstring theory acts as a physical
regulator for its limiting supergravity theory, so the orders at which various
operators become relevant to supergravity divergences gives information about
the presence of the same operator with a finite coefficient in superstring
theory.

Many problems of current interest involve the {\em scaling} behavior of quantum
corrections, which is determined by how strongly they are renormalized. And
the verification of duality conjectures makes detailed use of supergravity
radiative corrections. And a number of mysteries have arisen: some correlation
functions appear to be mysteriously ``protected'' for reasons that are not yet
understood. So, in this context it is important to get to the bottom of any
disagreements between different viewpoints on the ultraviolet problem.

One good way to approach the study of non-renormalization theorems is through the
background-field method. A classic example of a non-renormalization theorem is
the Adler-Bardeen theorem \cite{stelle:adlerbardeen}. This may be understood
from the standpoint of the background field method \cite{stelle:bms} as follows.
Consider the Yang-Mills Chern-Simons operator under a background-quantum split 
$A^r_a={\cal B}^r_a+Q^r_a$:
\be
K^a=4g^2\epsilon^{abcd}[A^r_b(\partial_cA^r_d+\ft13gf^{rst}A^s_cA^t_d)]\
.\label{stelle:eq1}
\ee
This operator has the property that its divergence gives the Pontryagin density:
$\partial_aK^a=2g^2F^r_{ab}{}^\ast F^{r\,ab}$, but it is not itself gauge
invariant. It's renormalization properties, however, are important in the
context of anomalies. When one expands it in powers of the quantum field
$Q^r_a$, the lowest term is just the expression (\ref{stelle:eq1}) with
${\cal B}^r_a$ substituted for $A^r_a$, so it is no more background gauge
invariant than (\ref{stelle:eq1}). Nor is the term linear in $Q^r_a$ invariant
under background gauge transformations. But the term quadratic in $Q^r_a$ can be
written $4g^2\epsilon^{abcd}Q_a^rD_b({\cal B})Q_c^r$, which is fully background
gauge invariant. Similarly, the term cubic in $Q^r_a$ is background invariant,
since $Q^r_a$ transforms as a tensor under the background transformations.

The above structure for $K^a$ has the striking consequence that although $K^a$
itself is not gauge invariant, the terms actually used in calculating
one-particle-irreducible (1PI) Feynman diagrams with operator insertions of
$K^a$ are actually background gauge invariant. Accordingly, the renormalization
of this operator can only be through gauge invariant operators, of which class
it is not itself a member. Consequently, neither $K^a$ itself nor does any
gauge-invariant operator
$O$ mix with it under renormalization, and so for the anomalous
dimensions one has $\gamma_{\ssst KK}=\gamma_{\ssst OK}$=0.

The Adler-Bardeen theorem then arises as follows. In the presence of axial
anomalies, the axial current $j^5_a$ is not conserved;
one has the famous relation $\partial_aJ^{5\,a}=2g^2cF^r_{ab}{}^\ast F^{r\,ab}$,
where $c$ is the anomaly coefficient, first occurring at the one-loop level.
However, there is a non-gauge-invariant current $j'^5_a=J^5_a-cK_a$ that is
conserved. Since conserved operators are not renormalized, one has the
renormalization group equation $\mu{\partial\over\partial\mu}j'^5_a=0$.
In consequence, and taking into account the $K^a$ anomalous dimensions from
above, $\gamma_{\ssst KK}=\gamma_{j{\ssst K}}=0$ together with independence of
the operators $j^5_a$ and $K_a$, one finds
\bea
\gamma_{jj}-c\gamma_{{\ssst K}j}&=&0\\
\mu{\partial c\over\partial\mu}&=&0\ ;\label{stelle:eq3}
\eea
and (\ref{stelle:eq3}) is just the Adler-Bardeen theorem. 

We may draw from this brief review the moral that ``vestigial'' consequences of
gauge invariance, such as those revealed by the above use of the background
field method, can give rise to important non-renormalization properties, such as
the Adler-Bardeen theorem.

Now let us review the basic non-renormalization theorem \cite{stelle:grs} of
$N=1$ supersymmetry, also from the point of view of the background field method.
We consider the basic Wess-Zumino model, based upon a chiral superfield $\phi$;
$\bar D_{\dot\alpha}\phi=0$. In the background field method, one again makes a
background-quantum split, $\phi=\varphi+Q$. The background superfield $\varphi$
then appears only on the external legs of 1PI Feynman diagrams used in
calculating the effective action $\Gamma(\varphi)$, while the quantum field $Q$
occurs on the inside lines of diagrams. 

The derivation of the Feynman rules for the $Q$ lines may be carried out either
taking particular care with the functional delta functions arising from
variational derivatives with respect to chiral superfields, or one may solve the
chirality constraint for $Q$ and work henceforth with unconstrained general
scalar superfields. Adopting the second approach, one writes $Q=\bar D^2 X$,
where $\bar D^2=\bar D^{\dot\alpha}\bar D_{\dot\alpha}$. Expanding then the
action into background and quantum fields, one finds that all terms except for
the ${\cal O}(X^0)$ terms can now be re-written as integrals over the full
superspace. Thus, \eg, a mass term
decomposes as $m\int d^4xd^2\theta\phi^2\longrightarrow
m\int d^4xd^2\theta\varphi^2+2m\int d^4xd^4\theta\varphi X+\int d^4xd^4\theta
X\bar D^2X$.

The vertices and propagators for the quantum $X$ fields used in 1PI diagrams are
derived from terms containing 2 or more $Q$ fields, expanded as above and
written as full superspace integrals. On the other hand, the background
$\varphi$ superfields continue to appear always as constrained, chiral,
superfields. Hence the non-renormalization theorem for $D=4$, $N=1$
supersymmetry arises: perfectly supersymmetric invariants such as chiral
superspace integrals over prepotentials may occur in the original action of
the theory, but they may not occur as counterterms. From the structure of the
Feynman rules, all quantum corrections to the effective action $\Gamma$ must be
written as full superspace integrals. Thus, chiral superspace integrals like
$\int d^4x d^2\theta f(\phi)$ are not renormalized. Note that one cannot try to
frustrate the strictures of this non-renormalization theorem by writing chiral
superspace integrals in forms like $\int d^4x d^4\theta
\phi\square^{-1}D^2\phi$, because the operator $\square^{-1}$ is spatially
non-local, and this then is not in accord with the requirements of Weinberg's
theorem \cite{stelle:weinberg} on the locality of counterterms.

For extended supersymmetries that can be given a complete linear realization in
superspace, similar results are found. In particular, for matter fields such as
the $N=2$ hypermultiplet, one directly obtains a similar result: hypermultiplet
self-interactions are not renormalized \cite{stelle:hst1}. 

For gauge multiplets and supergravity multiplets, there is a further subtlety,
which affects however only the one-loop diagrams. As in the case of the
$N=1$ chiral multiplet, one introduces prepotentials for the quantum fields that
appear on internal lines of diagrams, while background fields appear through
constrained superfields, the original superspace gauge connections.
The procedure of gauge fixing and introduction of the ghost actions, however,
introduces some complications. Superspace gauge symmetry
parameters are described by constrained superfields, \eg\ the chiral gauge
superfield $\Lambda$ for $N=1$, $D=4$ super Yang-Mills theory, which is
constrained in the background field method by a background-covariant constraint
$\bar{\cal D}_{\dot\alpha}({\cal B})\Lambda=0$. In the construction of the
corresponding ghost action for a chiral ghost superfield written with the
chirality constraint solved {\it via} the introduction of a prepotential, \eg\ 
$\bar{\cal D}^2 S$, one finds that $S$ has its own gauge invariance $\delta
S=\bar{\cal D}^{\dot\alpha}\bar\Lambda_{\dot\alpha}$. This requires a new
gauge fixing and ghost action, which in turn has a new gauge invariance with
chiral parameter $\bar\Lambda_{\dot\alpha\dot\beta}$. This process continues
indefinitely, with further gauge parameters
$\bar\Lambda_{\dot\alpha\dot\beta\dot\gamma}$,
\etc, and these will all couple to the background fields unless one cuts off the
interactions with the background at some order. This necessitates the
introduction of a background gauge field prepotential, thus breaking the
non-renormalization theorem, but only for the one-loop diagrams.

\section{Divergence estimates for extended
supersymmetry}

Theories with extended supersymmetry, and in particular the maximal supergravity
and super Yang-Mills theories, face another problem in estimating the onset of
ultraviolet divergences: not all of a given theory's supersymmetry can be
linearly realized ``off-shell,'' \ie\ with an algebra that closes without use
of the equations of motion. This is an old problem in supersymmetric theories,
whose full implications are still to this day not clear. In analyzing the
possibilities for infinities, one can choose between several alternatives for
the ``next best thing.'' For example, one can see what is the maximal degree of
supersymmetry that can be linearly realized off-shell and then use that, keeping
at the same time full Lorentz invariance \cite{stelle:hst2}. Or, alternatively,
one can keep the full automorphism symmetry that appears in the on-shell
formalism without auxiliary fields but sacrifice Lorentz invariance, \eg\ 
through use of a light-cone formalism \cite{stelle:mandelstam-bln}. Similar
conclusions are obtained in either formalism; however, the Lorentz-covariant
approach has been used to study more cases. The upshot is that for
$N=4\leftrightarrow 3,2,1$ super Yang-Mills one can use formalisms with
$M=2,2,1$ off-shell supersymmetry. For
$N=8\leftrightarrow7,6,5,4,3,2,1$ supergravity theories one can use formalisms
with $M=4,3,3,2,2,2,1$ off-shell supersymmetry.

Putting all of this together, one has a general supersymmetric
non-renormalization theorem:
{\it For gauge and supergravity multiplets at loops $\ell\ge2$, and for matter
multiplets at all loop orders, counterterms must be written as full superspace
integrals for the maximal off-shell linearly realizable supersymmetry, with
background gauge invariant integrands written without using prepotentials for the
background fields.}

Although the full power of a given theory's supersymmetry cannot always be used
to determine the structure of counterterms from supersymmetric Ward identities
-- only the linearly realized supersymmetry gives useable Ward identities in
general -- the full supersymmetry can still play a r\^ole. Things simplify
greatly if one considers the counterterms subject to the classical field
equations. For the first non-vanishing one of these, the Ward identities for the
full (nonlinear) supersymmetry requires full supersymmetry invariance, subject
to the imposition of the classical field equations \cite{stelle:hs}. There is no
requirement, however, on being able to write the counterterm as a full
superspace integral for the full nonlinear supersymmetry. Nonetheless, this
requirement does serve to rule out some counterterms that would by themselves be
acceptable according to the above theorem, but whose coefficients become linked
{\it via} the full supersymmetry to the coefficients of terms disallowed by the
above theorem.

Of course, the above considerations cannot say exactly when the first
divergences can occur in a given theory -- the most they can say is up to which
loop orders the divergences cannot. Nonetheless, in the study of divergences one
has become used to the maxim that if something isn't ruled out, it will occur.
So any further cancellations would appear ``miraculous'' from the standpoint of
the above analysis. Here is a table of the expectations for divergences in
maximal (\ie\ $N=1$ in $D=10$ or $N=4$ in $D=4$) super Yang-Mills theory:
\begin{table}[ht]
\centering
\caption{Super Yang-Mills divergence expectations, standard
Feynman rules.\label{stelle:tab1}}
\begin{tabular}{|l|c|c|c|c|c|c|}
\phantomthreeptline\nhline
Dimension&10&8&7&6&5&4\\
loop $L$&1&1&2&3&4&$\infty$\\
gen.\ form&$\partial^2F^4$&$F^4$&$\partial^2F^4$&$\partial^2F^4$&$F^4$&finite\\
\nhline
\nhline\phantomthreeptline
\end{tabular}
\end{table}
\newline and in maximal (\ie\ $N=1$ in $D=11$ or $N=8$ in $D=4$) supergravity:
\begin{table}[ht]
\centering
\caption{Supergravity divergence expectations, standard
Feynman rules.\label{stelle:tab2}}
\begin{tabular}{|l|c|c|c|c|c|c|c|}
\phantomthreeptline\nhline
Dimension&11&10&8&7&6&5&4\\
loop $L$&2&1&1&2&2&2&3\\
gen.\
form&$\partial^6R^4$&$\partial^2R^4$&$R^4$&$\partial^4R^4$&$\partial^2R^4$&$R^4$&$R^4$\\
\nhline
\nhline\phantomthreeptline
\end{tabular}
\end{table}

\section{String-inspired calculations via cutting rules}

Traditional Feynman diagram techniques quickly become unmanageable. For example,
a 5-loop diagram in $N=8$, $D=4$ supergravity may have on the order of $10^{30}$
terms in its algebraic expression. So a better procedure is clearly desirable.

Just such a procedure (let us call it the BD$^2$PR procedure
\cite{stelle:bddpr}) has been offered in recent years, using factorization
properties of supergravity amplitudes inherited from string theory, coupled with
a procedure for extracting the ultraviolet divergences from the absorptive parts
of amplitudes, using the cutting rules. The key elements of this procedure are:
\begin{enumerate}
\item[1)] the optical theorem
\item[2)] dimensional regularization
\end{enumerate}

Specifically, from the optical theorem, one determines the imaginary part of a
one-loop amplitude in terms of a product of tree amplitudes, for example. One
might worry that by this procedure one looses information about the real parts
of amplitudes. But this is avoided by a careful use of dimensional
regularization. The optical theorem works best for amplitudes containing
logarithms, where one can clearly see the relation between real and imaginary
parts, \eg\ $\ln(s)=\ln(|s|)+\im\pi\theta(s)$. This can be enhanced by
dimensional regularization, which produces many additional logarithms, \eg\
$s^{-2\epsilon}=1-2\epsilon\ln(s)+\ldots$, where $\epsilon=D-D_{\rm physical}$;
these help to determine amplitudes providing one calculates diagrams factorized
using the cutting rules {\em to all powers in $\epsilon$.} The remaining
essentials in the BD$^2$PR procedure are
\begin{enumerate}
\item[3)] field-theory amplitudes result from string amplitudes in the
$\alpha'\rightarrow 0$ limit
\item[4)] the KLT relations \cite{stelle:klt} between closed and open string
tree-level S-matrices.
\end{enumerate}

As a combined example of the two last points, one has a relation between
supergravity and super Yang-Mills 4-point amplitudes, arising from the
$\alpha'\rightarrow 0$ limit of superstring theory, where $M_4^{\rm tree}$ is a
supergravity amplitude and $A_4^{\rm tree}$ is a SYM amplitude:
\be
M_4^{\rm tree}(1,2,3,4) = -\im s_{12} A_4^{\rm tree}(1,2,3,4)A_4^{\rm
tree}(1,2,3,4)\ .
\ee

\begin{figure}[ht]                                                
  {\includegraphics[height=.3\textheight]{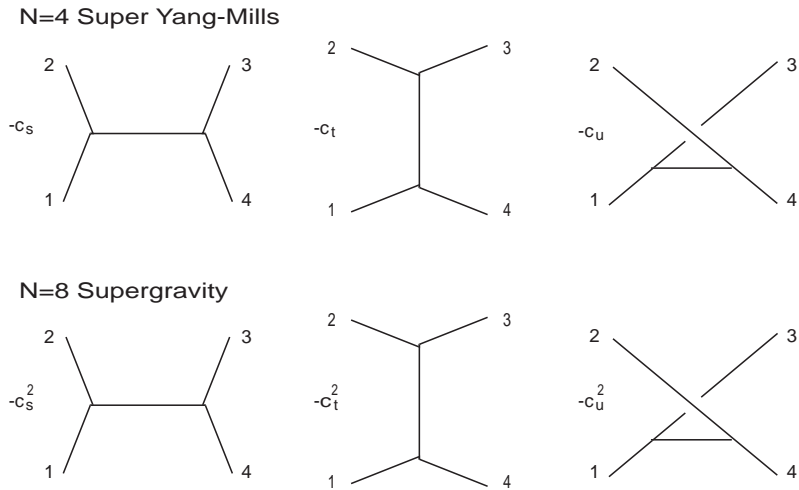}}  
  \caption{Tree-level color-ordered $N=4$ SYM and $N=8$
SG amplitudes expressed in terms of $\phi^3$
diagrams. Supergravity coefficients are squares of the
SYM coefficients.\label{stelle:fig1}}
\end{figure}

Using the KLT relations, one may organize supergravity amplitudes and super
Yang-Mills amplitudes to display a squaring relationship. Moreover, amplitude
calculations with irreducible supermultiplets reduce to scalar amplitudes
multiplied by index-bearing factors. This is illustrated in Fig.\
\ref{stelle:fig1}.

\begin{figure}[ht]                                            
  {\includegraphics[height=.15\textheight]{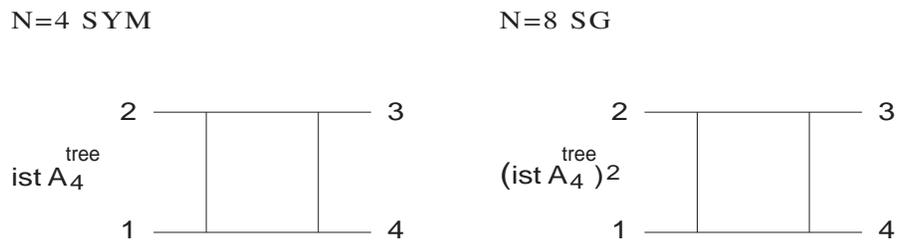}}  
  \caption{Relation between $N=4$ SYM theory and $N=8$ SG amplitudes at the
one-loop level. Tree amplitudes multiply scalar field loop diagrams.
\label{stelle:fig2}}
\end{figure}

The relations between supergravity and super Yang-Mills amplitudes continues on
to higher loops, enabling SG and SYM amplitudes to be calculated in terms of
scalare diagrams. A one-loop example is illustrated in Fig.\ \ref{stelle:fig2}

Using these techniques, BD$^2$PR obtain the following predictions for the first
on-shell divergences in the maximal super Yang Mills theories:

\begin{table}[ht]
\centering
\caption{Maximal super Yang-Mills divergence expectations from
cutting rules.\label{stelle:tab3}}
\begin{tabular}{|l|c|c|c|c|c|c|}
\phantomthreeptline\nhline
Dimension&10&8&7&6&5&4\\
loop $L$&1&1&2&3&6&$\infty$\\
gen.\
form&$\partial^2F^4$&$F^4$&$\partial^2F^4$&$\partial^2F^4$&
$\partial^2F^4$&finite\\
\nhline
\nhline\phantomthreeptline
\end{tabular}
\end{table}
\noindent and in maximal supergravity they obtain:
\begin{table}[ht]
\centering
\caption{Maximal supergravity divergence expectations from cutting
rules.\label{stelle:tab4}}
\begin{tabular}{|l|c|c|c|c|c|c|c|}
\phantomthreeptline\nhline
Dimension&11&10&8&7&6&5&4\\
loop $L$&2&1&1&2&3&4&5\\
gen.\
form&$\partial^6R^4$&$\partial^2R^4$&$R^4$&$\partial^4R^4$&$\partial^6R^4$&
$\partial^6R^4$&$\partial^4R^4$\\
\nhline
\nhline\phantomthreeptline
\end{tabular}
\end{table}

\newpage
So one can see that the BD$^2$PR results appear to be predicting a ``miracle''
in the onset of divergences in the maximal SYM and SG theories. One may
summarize their results for the two maximal theories by the rule that the
maximal SYM theories are finite for $D<{6\over L}+4$ ($L>1$) and the maximal
supergravity theories are finite for $D<{10\over L}+2$ ($L>1$). A striking
aspect of this improvement is that it applies only to the maximal supersymmetric
theories. The cutting-rule analysis of $N\le 6$ supergravity in $D=4$ shows the
onset of divergences to be exactly as predicted on the basis of standard Feynman
diagram analysis, equal to those summarized for the maximal theories in tables
\ref{stelle:tab1} and
\ref{stelle:tab2}.

Of course, there is strictly speaking no firm contradiction here, for the
standard Feynman diagram analysis gave only a lower bound on the possible loop
order for the first on shell surviving infinities. Nonetheless, the difference
is striking. The BD$^2$PR results are in fact more than an estimate. They give
an actual calculation of the coefficients of the various divergences, limited
only in that for loops $L>2$ only two-particle cuts have been considered; a full
calculation would require general $m$ particle cuts.

\section{Counterterm analysis}

The study of counterterms for super Yang-Mills and supergravity theories has
been going on since the discovery of these theories. Pure Einstein gravity
diverges at the two-loop order with an $R^3$ counterterm, as found by explicit
calculations \cite{stelle:gsvdv}. In supergravity theories, on-shell
supersymmetry rules out all on-shell surviving counterterms at the $L=1$ and
$L=2$ loop orders. In particular, this dramatically causes the cancellation of
the $R^3$ counterterm that is present in pure Einstein gravity.

The first dangerous supergravity counterterm occurs at $L=3$ loops in $D=4$, and
has an $R^4$ generic structure. This counterterm involves the curvature through
the square of the Bel-Robinson tensor
\be
T_{\mu\nu\rho\sigma}=R_{\mu\alpha\nu\beta}R^\alpha{}_\rho{}^\beta{}_\sigma +
{}^\ast R_{\mu\alpha\nu\beta}{}^\ast R^\alpha{}_\rho{}^\beta{}_\sigma\ .
\ee
This tensor is a kind of analogue of the stress tensor for gravity: it is
totally symmetric and totally traceless (for pure gravity) and it is covariantly
conserved on any index. In an analogy to the supersymmetric
$T_{\mu\nu}T^{\mu\nu}$ counterterms that exist for matter systems in a
gravitational background, pure supergravity has a
$T_{\mu\nu\rho\sigma}T^{\mu\nu\rho\sigma}+{\rm fermionic}$ counterterm
\cite{stelle:dks}, appropriate by standard power counting for the $L=3$ order of
perturbation theory. The existence of similar counterterms was also found for
all $D=4$ extended supergravities \cite{stelle:dk,stelle:kallosh,stelle:hst3}.
These same expressions occur as important corrections with finite coefficients to
superstring theories; the question in that case is also at which loop order they
first appear.

Consider now in some more detail the structure of the $R^4$ counterterm for the
higher $N\ge4$ supergravities in $D=4$ \cite{stelle:hst3}. The on-shell
supergravity multiplet is described by a superfield $W_{[ijkl]}$ carrying
automorphism group ${\rm SU}(N)$ indices and satisfying the full on-shell
superspace constraints
\bea
D_\alpha^iW_{jklm}&=&{-4\over N-3}\delta^i_{[j}D_\alpha^nW_{klm]n}\\
\bar D_{\dot\alpha\,i}W_{jklm}&=&\bar D_{\dot\alpha\,[i}W_{jklm]}\
.\label{stelle:wconstraints}
\eea
At the linearized level, which all that we need concern ourselves with here,
each component of $W_{ijkl}$ is a field $F$ with $2s$ symmetrized spinor indices
of the same chirality. The constraints imply
$\partial^{\alpha_1}_{\dot\alpha}F_{\alpha_1\cdots\alpha_{2s}}=\square
F_{\alpha_1\cdots\alpha_{2s}}=0$. These on-shell component field strengths
describe massless spin $s$ for $s\le2$, including $N(N-1)(N-2)(N-3)/12$ $s=0$
scalars for $N\le7$, and $70$ scalars for the self-conjugate $N=8$ theory. All
of these scalars arise from dimensional reduction of the metric and the
higher-dimensional gauge field $A_{\mu\nu\rho}$.

The $R^3$ counterterm is written for the $N=8$ theory as an on-shell superspace
integral over a subset of 16 out of the 32 fermionic coordinates. We recall that
the non-renormalization theorem requires that this counterterm (which is the
first one non-vanishing on shell) be fully supersymmetric under the full (in
general nonlinear, although here we consider only the linearized level) $N=8$
supersymmetry, but that it also be possible to be expressed in off-shell $N=4$
superfields in which no prepotentials are introduced and the counterterm is
written as a full superspace integral. The $\int d^{16}\theta$ structure of the
$R^4$ counterterm allows this to happen. In linearized superspace, the structure
of the on-shell counterterm is \cite{stelle:hst3}
\be
\Delta\Gamma=\kappa^4\int d^4x(d^{16}\theta)_{232848}(W^4)_{232848}
\ee
where the 232848 representation of ${\rm SU}(8)$ is described by the
$4\times 4$ square Young tableau with four rows of four boxes each. One can
also view this as a superinvariant built from the 232848 rep.\ extracted from
the quadratic product of two $(W^2)_{1764}$ superfields, where the 1764 rep.\ 
is described by two columns of 4 boxes. This is the multiplet that contains the
Bel-Robinson tensor, which in 2-component indices is
$T_{\alpha\dot\alpha,\beta\dot\beta,\gamma\dot\gamma,\delta\dot\delta}
=C_{\alpha\beta\gamma\delta}C_{\dot\alpha\dot\beta\dot\gamma\dot\delta}$, where
$C_{\alpha\beta\gamma\delta}$ is the Weyl tensor in two-component notation.
Since this counterterm satisfies all the requirements of the non-renormalization
theorem discussed earlier.

\section{Gauge invariance and the discrepancy in divergence
estimates}

It is striking that the standard Feynman diagram analysis and the BD$^2$PR
analysis agree in almost all cases, with the only disagreements coming for the
maximal super Yang-Mills and supergravity theories. This provides a clue to what
might be going on. Again, it seems to come down to a difference in the treatment
of gauge invariance. Both of the theories for which there is a discrepancy have
the property that they can be dimensionally oxidized to theories in which all
the scalar fields have disappeared into components of higher-dimensional gauge
fields, metrics, and so on. In particular, the $D=4$, $N=8$ supergravity theory
oxidizes to $D=11$, $N=1$ supergravity, in which {\em all} fields are gauge
fields, including all the spinors.

This observation implies that the origin of the difference in divergence
estimates lies in the way gauge invariance is being treated. Indeed, if one
adopts the rule that background fields must appear in the counterterm in a way
that can also be oxidized to a gauge-invariant structure in the higher
dimension, then the two divergence estimates agree \cite{stelle:hps}. This also
works for the super Yang-Mills theories. 

Another example that shows how the discrepancy between the two divergence
estimates is removed by the revised rule for gauge invariance occurs in $D=5$
maximal super Yang-Mills theory. This has 5 scalars, but all of them are
absorbed upon oxidization of the theory to $D=10$, $N=1$ super Yang-Mills
theory. Another example is $D=5$, $N=1$ SYM, which has a single scalar. This
can, however, be absorbed into the gauge field upon oxidization to $D=6$, $N=1$
SYM theory.

This leaves us with a question, on which we are going to end this review of
divergence problems and their attendant mysteries. {\it Are there circumstances
in which higher-dimensional gauge symmetries may still effectively govern the
structure of counterterms in lower-dimensional theories?} If so, then the
standard superspace Feynman rules and the BD$^2$PR cutting-rule analysis give
the same results. But how this comes about remains mysterious. There does not
appear to be any direct way to preserve more of the higher-dimensional gauge
symmetries than that which is manifest in the lower dimension upon dimensional
reduction. The gauge transformations depending upon the compactified
dimensions are the ones that would have to impart structure to the
counterterms, but these are lost when one suppresses dependence upon the
compactified coordinates, or sets to zero the corresponding loop momenta.

So we end with this question: do supersymmetry and the vestiges of superstring
theory present in supersymmetric gauge theories conspire to keep alive the
higher gauge symmetries? These are the most unified theories that we know, and
it just might be that in whatever dimension one formulates them, they remember
their higher-dimensional origins.

\end{document}